\documentclass[conference]{IEEEtran}
\usepackage{graphicx,cite,bm,algorithm,algorithmic,psfrag,amsmath}

\def\argmax{\mathop{\mbox{arg\,max}}}
\newcommand{\defequal}{\stackrel{\mathrm{def}}{=}}
\renewcommand{\vec}[1]{{\ensuremath{\boldsymbol{#1}}}}
\newcommand{\popt}{\ensuremath{P^{(K)}_{opt}}}
\IEEEoverridecommandlockouts
\begin{document}
\title{Opportunistic Collaborative Beamforming with One-Bit Feedback}
\author{Man-On Pun, D.~Richard Brown III and H.~Vincent Poor
\thanks{Man-On Pun and H.~Vincent Poor are with the Department of Electrical Engineering, Princeton University, Princeton, NJ 08544 (e-mail: mopun@princeton.edu; poor@princeton.edu).}
\thanks{D.~Richard Brown III is visiting Princeton University from the Electrical and Computer Engineering Department, Worcester Polytechnic Institute, Worcester, MA 01609. (e-mail: drb@wpi.edu).}
\thanks{This research was supported in part by the Croucher Foundation under a post-doctoral fellowship, and in part by the U.S.~National Science Foundation under Grants ANI-03-38807, CNS-06-25637, and CCF-0447743.}}
\maketitle
\maketitle \thispagestyle{plain}
\begin{abstract}
An energy-efficient opportunistic collaborative beamformer with one-bit feedback is proposed for ad hoc sensor networks over Rayleigh fading channels. In contrast to conventional collaborative beamforming schemes in which each source node uses channel state information to correct its local carrier offset and channel phase, the proposed beamforming scheme opportunistically selects a subset of source nodes whose received signals combine in a quasi-coherent manner at the intended receiver. No local phase-precompensation is performed by the nodes in the opportunistic collaborative beamformer. As a result, each node requires only one-bit of feedback from the destination in order to determine if it should or shouldn't participate in the collaborative beamformer. Theoretical analysis shows that the received signal power obtained with the proposed beamforming scheme scales {\em linearly} with the number of available source nodes. Since the the optimal node selection rule requires an exhaustive search over all possible subsets of source nodes, two low-complexity selection algorithms are developed. Simulation results confirm the effectiveness of opportunistic collaborative beamforming with the low-complexity selection algorithms.
\end{abstract}

\section{Introduction}
Collaborative beamforming has recently attracted considerable research attention as an energy-efficient technique to exploit distributed spatial diversity in ad hoc sensor networks \cite{Ochiai05, Mudumbai07,Brown05}. In collaborative beamforming, a cluster of low-cost and power-constrained source nodes collaboratively transmit a common message to a distant destination node, e.g.~a base station (BS) or an unmanned aerial vehicle. It has been demonstrated that collaborative beamforming can provide substantially improved data rate and transmission range by forming a {\em virtual} antenna array to direct transmitted signals towards the destination node \cite{Ochiai05, Mudumbai07}. However, similar to the conventional beamforming techniques, collaborative beamforming requires perfect channel state information (CSI) at each source node in order to achieve coherent combining at the intended destination. More specifically, each source node must pre-compensate its any local carrier offset as well as any phase distortion caused by its channel such that the bandpass signals from all the nodes arrive at the receiver with identical phase. Without properly adjusting the phases of transmitted signals, collaborative beamforming may perform poorly due to pointing errors and mainbeam degradation \cite{Ochiai05}.

To obtain CSI, the source nodes can exploit pilot signals transmitted from the BS by assuming channel reciprocity. However, since this approach involves channel estimation at each source node, it imposes hardware penalties on the systems, which is undesirable for developing low-cost networks. Alternatively, CSI can be estimated by the BS and returned to the source nodes. While this approach allows for low-complexity source node hardware, it may incur excessive feedback overhead, particularly for networks comprised of a large number of source nodes. To circumvent this problem, two novel approaches have been developed in the literature. In \cite{Madan06}, only a subset of the available source nodes with the largest channel gains are selected for collaborative beamforming. As a result, the total amount of CSI feedback is reduced proportionally to the number of selected source nodes. Accurate phase feedback, however, may still require many bits of information per selected node. By contrast, feedback is completely eliminated in \cite{Bletsas06} where a distributed scheme was proposed to select the single source node with the strongest channel gain. This approach, however, eliminates feedback by sacrificing the potential beamforming gains.

In this work, we propose {\em opportunistic} collaborative beamforming with one-bit feedback. Inspired by the observation that bandpass signals with even moderate phase offsets can still combine to provide beamforming gain,
the proposed scheme opportunistically selects a subset of available source nodes whose transmitted signals combine in a quasi-constructive manner at the intended receiver. Unlike conventional collaborative beamforming, no local phase-precompensation is performed by the source nodes. As a result, each node requires only one-bit of feedback from the destination in order to determine if it should or shouldn't participate in the collaborative beamformer. Theoretical analysis shows that the received signal power obtained with the proposed beamforming scheme scales linearly with the number of available source nodes. Since the the optimal node selection rule is exponentially complex in the number of available nodes, two low-complexity selection algorithms are developed. Simulation results confirm the effectiveness of opportunistic collaborative beamforming with the low-complexity selection algorithms.

\underline{Notation}: Vectors and matrices are denoted by boldface letters. Furthermore, we use $E\left\{\cdot\right\}$, $\left(\cdot\right)^T$ and $\left(\cdot\right)^H$ for expectation, transposition and Hermitian transposition.

%\section{Description of Proposed Scheme}
%We assume that the channel between each source node and the BS undergoes independent Rayleigh fading with a channel coherence time sufficiently larger than the symbol duration. As a result, the channels are approximately constant over a few data symbols. Adopting a scheme similar to the channel sounding mechanism standardized in IEEE 802.16e \cite{80216e}, the BS can easily estimate the CSI of each source node over the sound zone. After selecting the source nodes based on the proposed selection algorithms, the BS broadcasts a short binary selection message to the source nodes. If a source node is selected, it will proceed to data transmission. Otherwise, the source node waits for the next channel sounding. To concentrate our study on the selection scheme design, we assume that perfect synchronization and channel estimation have been achieved in this work.

\section{Signal Model}
%%%%%%%%%%%%%%%%%%%%%%%%%%%%%%%%%%%%%%%%%%%%%%%%%%%%%%%%%%%%%%%%
\begin{figure}[htp]
\begin{center}
\includegraphics[scale=1]{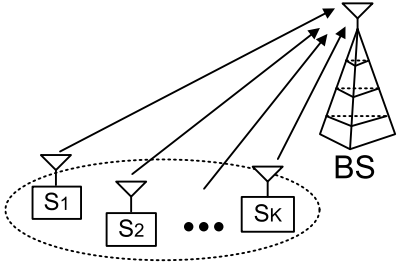}
\caption{System under consideration for collaborative beamforming.}\label{fig:coop}
\end{center}
\end{figure}
%%%%%%%%%%%%%%%%%%%%%%%%%%%%%%%%%%%%%%%%%%%%%%%%%%%%%%%%%%%%%%%%
We consider a single-antenna network comprised of $K$ source nodes and one destination node (the BS) as illustrated in Fig. \ref{fig:coop}. The channel gain between the $k$-th source node and the BS, denoted by $h_k$, is modeled as ${\cal CN}\left(0,1\right)$ with
\begin{equation}
h_k=a_ke^{j\phi_k}, \quad k=1,2,\cdots,K.
\end{equation}
where $a_k\geq 0$ and $\phi_k\in\left(-\pi,\pi\right]$ are the Rayleigh-distributed channel amplitude and uniformly-distributed channel phase, respectively. Furthermore, $a_k$ and $\phi_k$ are assumed statistically independent of each other over all source nodes.

Denote by ${\bm s}$ the selection vector of length $K$. The $k$-th entry of ${\bm s}$ is one, i.e. $s_k=1$, if and only if the $k$-th source node is selected for transmission; otherwise $s_k=0$. Thus, the received signal can be written as
\begin{equation}\label{eq:rxsignal}
r=\frac{1}{\sqrt{{\bm s}^T{\bm s}}}{\bm h}^T{\bm s}d+v,
\end{equation}
where $d$ is the unit-power data symbol, ${\bm h}=\left[h_1,h_2,\cdots,h_K\right]^T$ and $v$ is complex Gaussian noise modeled as ${\cal CN}\left(0,\sigma^2\right)$. It should be emphasized that the total transmitted signal power is normalized to unity, regardless the number of selected source nodes. As a result, a collaborative beamforming scheme is more energy efficient if it provides a higher received signal power than single-source transmission.

\section{Two-Node Beamforming}
\label{sec:twonode}
To shed light on the beamforming gain of the proposed scheme, we first consider the case when two source nodes are available for cooperative transmission. We assume without loss of generality that $a_1 \ge a_2$. Then we can say
\begin{eqnarray}
  P_{\{1\}} = a_1^2 \ge a_2^2 = P_{\{2\}}.
\end{eqnarray}
When both sources transmit, the received power can be expressed as
\begin{eqnarray}
  P_{\{1,2\}} &=& \frac{1}{2} \left| a_1 e^{j \phi_1}+a_2 e^{j \phi_2}\right|^2, \\
           &=& \frac{a_1^2}{2} \left| 1+\rho e^{j \Delta}\right|^2,
\end{eqnarray}
where $\rho \defequal a_2/a_1$ and $\Delta \defequal\phi_2-\phi_1$. Simultaneous transmission is optimal if $P_{\{1,2\}} \ge P_{\{1\}}$, which corresponds to the equivalent condition
\begin{eqnarray}
  \label{eq:suff2nodes}
  \cos(\Delta) & \ge &  \frac{1-\rho^2}{2 \rho}.
\end{eqnarray}
The following special cases of (\ref{eq:suff2nodes}) are of interest.
\begin{itemize}
\item When $\rho=1$, both sources have identical channel amplitudes and the simultaneous transmission condition in (\ref{eq:suff2nodes}) reduces to $|\Delta| \le \frac{\pi}{2}$. The gain with respect to single-source transmission, the case considered in \cite{Bletsas06}, can be expressed as
\begin{eqnarray}
  \Gamma = \frac{P_{\{1,2\}}}{P_{\{1\}}} = \frac{1}{2}\left| 1+ e^{j \Delta}\right|^2,
\end{eqnarray}
which attains a maximum value of 2 when $\Delta = 0$ and a minimum value of 1 when $\Delta = \pm \frac{\pi}{2}$. Even relatively large phase offsets between the sources can lead to significant gains with respect to single-source transmission. For example, when $\Delta=\frac{\pi}{3}$, the resulting gain can be computed to be $\Gamma = 1.76$dB.
\item When $\Delta=0$, the transmissions from both sources arrive in perfect phase alignment at the destination. Interestingly, (\ref{eq:suff2nodes}) implies that simultaneous transmission is optimal only if $\rho \ge \sqrt{2}-1\approx 0.4142$. In other words, even though both nodes have perfect phase alignment, simultaneous transmission is optimal only if the ratio of the second node's channel amplitude to that of the first node is at least $0.4142$. \end{itemize}

\section{$K$-Node Beamforming}

The received power of a $K$-node opportunistic collaborative beamformer with the optimal selection rule can be written as
\begin{eqnarray}
\popt = \max_{\vec{s} \in \{0,1\}^K} \frac{1}{\vec{s}^T \vec{s}} |\vec{h}^T \vec{s}|^2.\label{eq:recpower}
\end{eqnarray}
Optimal selection of nodes that participate in the beamformer entails an exhaustive search over all possible $2^K-1$ possible selection vectors. As a result, the computational complexity required to obtain the optimal selection is formidable, even for a moderate value of $K$. To better understand the performance of the optimal opportunistic collaborative beamformer, this section develops lower and upper bounds on its performance for the large-network case, i.e.~$K \rightarrow \infty$. For finite $K$, we also propose an iterative greedy algorithm for source selection that adds one new source node in each iteration such that the resulting received power increases in each iteration.

\subsection{Large-Network Received Power Bounds}\label{sec:threshold}
Exploiting the inequality $|\vec{h}^T \vec{s}|^2\leq |\vec{a}^T \vec{s}|^2$ in (\ref{eq:recpower}), where ${\bm a}=\left[a_1,a_2,\cdots,a_K\right]^T$, an upper bound for $\popt$ can be derived by considering the case when all of the transmissions are received coherently at zero phase, i.e.~$h_k = a_k \ge 0$ for all $k\in\{1,\dots,K\}$. As discussed in Section~\ref{sec:twonode}, even though the nodes all combine constructively at the destination, the optimal beamforming selection rule should not select all $K$ nodes for simultaneous transmission. Instead, only nodes with sufficiently large amplitude should be selected such that the resulting {\em normalized} received power is maximized. Denoting the selection threshold as $r$, we can write
\begin{eqnarray}
  s_k &=& \begin{cases}
    1 & \text{ if } a_k \ge r \\
    0 & \text{ otherwise}.
  \end{cases}
\end{eqnarray}

Recall that $a_k$ are i.i.d.~Rayleigh distributed channel amplitudes with mean $\text{E}[a_k]=\frac{\sqrt{\pi}}{2}$. For sufficiently large $K$, we can say that
\begin{eqnarray}
\lim_{K\rightarrow\infty}\frac{\vec{s}^T\vec{s}}{K}= \Pr\left(a_k \ge r\right) = e^{-r^2}.
\end{eqnarray}
Thus, we can express the received power upper bound normalized by $K$ as
\begin{eqnarray}
  \lim_{K\rightarrow\infty}\frac{P^{(K)}_{ub}(r)}{K} &=& \lim_{K\rightarrow\infty}\frac{K}{\vec{s}^T\vec{s}}
  \left[  \int_{r}^\infty  2x^2 e^{-x^2}  \, dx  \right]^2 \\
  &=& \frac{\pi }{4} f(r),\label{eq:ubdexpr}
 \end{eqnarray}
where
\begin{eqnarray}
  \label{eq:fq}
  f(r) &\defequal& e^{r^2}\left[\text{erfc}(r)+\frac{2r}{\sqrt{\pi}}e^{-r^2} \right]^2,
\end{eqnarray}
with $\text{erfc}(x)$ being the complementary error function defined as $\text{erfc}\left(x\right)=\frac{2}{\sqrt{\pi}}\int_x^\infty e^{-t^2}\,dt$. Note that received power upper bound grows linearly with $K$, as would be expected of an ideal coherent beamformer. Numerical maximization of $f(r)$ can be performed to show that $\max f(r) \approx 1.0849$ and $r^* = \arg \max f(r) \approx 0.5316$. Hence, we can write
\begin{eqnarray}
  \label{eq:ub}
  \lim_{K\rightarrow\infty}\frac{\popt}{K} \le \lim_{K\rightarrow\infty}\frac{P^{(K)}_{ub}(r^*)}{K} = 0.8521.\label{eq:ubd}
\end{eqnarray}

%%%%%%%%%%%%%%%%%%%%%%%%%%%%%%%%%%%%%%%%%%%%%%%%%%%%%%%%%%%%%%%%
\begin{figure}[htp]
\begin{center}
\psfrag{r}[][Bl][0.95]{$r$}
\psfrag{a}[][Bl][0.95]{$\alpha$}
\psfrag{x}[][Bl][0.95]{$\text{Re}(h_k)$}
\psfrag{y}[l][Bl][0.95]{$\text{Im}(h_k)$}
\includegraphics[scale=0.5]{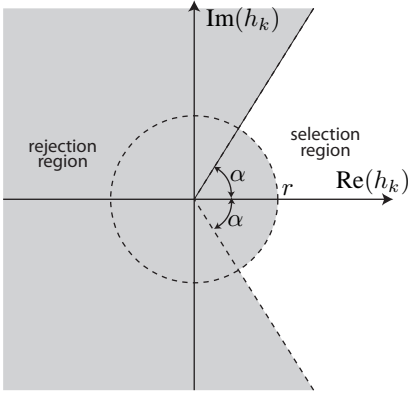}
\caption{Sector-based selection region used to derive the received power lower bound (\ref{eq:lb}).}\label{fig:selectionregion}
\end{center}
\end{figure}
%%%%%%%%%%%%%%%%%%%%%%%%%%%%%%%%%%%%%%%%%%%%%%%%%%%%%%%%%%%%%%%%

To develop a lower bound on $\popt$, we propose a suboptimal selection rule using the sector-based selection region shown in Fig.~\ref{fig:selectionregion}. The selection region is characterized by two parameters: $r$ corresponding to a minimum amplitude and $\alpha$ corresponding to a maximum angle. Nodes must satisfy both the minimum amplitude and maximum angle requirements to be selected for transmission, i.e.,
\begin{eqnarray}
  s_k &=& \begin{cases}
    1 & \text{ if } a_k \ge r \text{ and } |\phi_k| \le \alpha \\
    0 & \text{ otherwise}.
  \end{cases}
\end{eqnarray}

Given i.i.d.~channel coefficients $h_k = a_k e^{j \phi_k}$ with $a_k$ Rayleigh-distributed and $\phi_k$ uniformly distributed on $\left(-\pi,\pi\right]$, the probability that $h_k$ falls in the selection region $\Phi$ can be expressed as
\begin{eqnarray}
\Pr (h_k \in \Phi )&=&\Pr\left(|\phi_i|\leq \alpha\right)\Pr\left(a_i \ge r\right) \\
&=&\frac{\alpha}{\pi}\exp\left(-r^2\right).
\end{eqnarray}
When $K$ is large, the lower bound can be expressed as
\begin{eqnarray}
  \lim_{K\rightarrow\infty}\frac{P^{(K)}_{lb}(r,\alpha)}{K} &=&\lim_{K\rightarrow\infty} \frac{K}{\vec{s}^T \vec{s}}
  \left[
  \int_{-\alpha}^\alpha
  \int_{r}^\infty
   \frac{\cos \theta }{\pi}x^2 e^{-x^2}
  \, dx
  \, d\theta
  \right]^2 \nonumber\\
  &=& \frac{\sin^2\alpha}{4 \alpha } f(r),\label{eq:lbd}
 \end{eqnarray}
where we have used the fact that $\frac{\vec{s}^T\vec{s}}{K} \rightarrow \Pr\left(h_k \in \Phi\right)$ and $f(r)$ is as defined in (\ref{eq:fq}). The term $\frac{\sin^2\alpha}{4\alpha }$ is not a function of $r$ and attains its maximum when $\cos\alpha = \frac{\sin \alpha}{2\alpha}$. The optimum value $\alpha^* \approx 1.1656$ radians can be found numerically. Since $f(r)$ achieves its maximum at $r^* \approx 0.5316$, the received power lower bound can be written as
\begin{eqnarray}
  \label{eq:lb}
  \lim_{K\rightarrow\infty}\frac{P^{(K)}_{lb}(r^*,\alpha^*)}{K} =0.1965\le \lim_{K\rightarrow\infty}\frac{\popt}{K} \label{eq:lbdexp}
\end{eqnarray}
when $K$ is large. In the sequel, the selection algorithm employing $\left\{r^*,\alpha^* \right\}$ is referred to as the ``sector-based selection algorithm''.

Summarizing (\ref{eq:ubd}) and (\ref{eq:lbdexp}), the upper and lower bounds on the normalized received power of opportunistic collaborative beamforming with the optimum selection rule can be written as
\begin{eqnarray}
  \label{eq:bothbounds}
  0.1965 \le \lim_{K\rightarrow\infty}\frac{\popt}{K} \le 0.8521.
\end{eqnarray}
Two implications of this result merit further discussion:
\begin{enumerate}
  \item When $K$ is large, the ratio of the upper and lower bounds implies that $\popt$ will be no worse than 6.37dB below the power of the ideal coherent phase-aligned beamformer.
  \item When $K$ is large, even simple sub-optimal selection algorithms for opportunistic collaborative beamforming can result in a normalized received power that scales linearly with $K$. Since both the upper and lower power bounds are linear in $K$, the normalized received power of the optimum opportunistic collaborative beamformer must also scale linearly with $K$. This represents a significant improvement over the single-best-relay selection rule in \cite{Bletsas06} whose received power scales as $\log\left(K\right)$ \cite{Tse02}.
\end{enumerate}

\subsection{Iterative Greedy Selection Algorithm}
Despite its simplicity and insightful analytical results, the sector-based selection algorithm does not fully exploit the CSI available to the BS. In this section, an iterative greedy algorithm is proposed to select a sub-optimal subset of source nodes for collaborative beamforming with affordable computational complexity. Clearly, the success of the algorithm hinges on effectively determining the number of selected source nodes and identifying the suitable nodes. The proposed iterative algorithm successfully addresses these two issues by capitalizing on our previous analysis on the two-node case. In each iteration, the proposed algorithm adds one new node to the selection subset based on a well-defined cost function until no further beamforming gain can be achieved by adding more nodes.

We denote by $p^{(N)}\in\left\{1,2,\cdots,K\right\}$ the node index chosen in the $N$-th iteration, $1\leq N \leq K$. To facilitate our subsequent derivation, we first define the following two quantities:
\begin{eqnarray}
z^{(N)}&=&\frac{1}{\sqrt{N}}\sum_{n=1}^Na_{p^{(n)}}e^{j\phi_{p^{(n)}}},\\
P^{(N)}&=&\left|z^{(N)}\right|^2,
\end{eqnarray}
where $z^{(N)}$ is the composite channel gain between the $N$ selected source nodes and the BS while $P^{(N)}$ is the corresponding received signal power.

Now, we consider $P^{(N+1)}$ by adding one new source node into the subset of selected source nodes.
\begin{eqnarray}
P^{(N+1)}&=&\frac{1}{N+1}\left|\sum_{n=1}^{N+1}a_{p^{(n)}}e^{j\phi_{p^{(n)}}}\right|^2,\\
&=&\frac{1}{N+1}\left|\sqrt{NP^{(N)}}+a_{p^{(N+1)}}e^{j\Delta_{N+1}}\right|^2,\nonumber\\\label{eq:PN1}
\end{eqnarray}
where $\Delta_{N+1}$ is the {\em relative} phase offset between the newly added channel gain and $z^{(N)}$.

Next, we can rewrite (\ref{eq:PN1}) as
\begin{eqnarray}
P^{(N+1)}&=&\frac{1}{N+1}\left[NP^{(N)}+a_{p^{(N+1)}}^2+\right.\nonumber\\
&&\left.2a_{p^{(N+1)}}\sqrt{NP^{(N)}}\cos\left(\Delta_{N+1}\right)\right],
\end{eqnarray}

Clearly, the condition $P^{(N+1)}>P^{(N)}$ has to hold in order to incorporate the $p^{(N+1)}$-th source node into the collaborative transmission. After straightforward mathematical manipulation, the condition can be equivalently rewritten as
\begin{equation}
\cos\left(\Delta_{N+1}\right)>\frac{P^{(N)}-a_{p^{(N+1)}}^2}{2a_{p^{(N+1)}}\sqrt{NP^{(N)}}}.
\end{equation}

Finally, we are ready to propose the following iterative greedy selection algorithm. Denote by $\cal{I}$ the node index set containing source nodes selected for collaborative beamforming. Furthermore, let $\bar{\cal{I}}$ be the complementary set of $\cal{I}$ over $\left\{1,2,\cdots,N\right\}$. The proposed greedy algorithm is summarized in Algorithm \ref{al:itr}.

\begin{algorithm}
\caption{Iterative greedy selection algorithm} \label{al:itr}
\begin{algorithmic}
\STATES Initialize $N=1$, ${\cal I}=\left\{1\right\}$, $\bar{\cal I}=\left\{2,3,\cdots,K\right\}$, $z^{(1)}=a_1e^{j\phi_1}$ and $P^{(1)}=a_1^2$;
\PROCEDURE
\FOR{$N = 1$ to $K$}
\STATE Find
$i^*=\displaystyle\argmax_{i\in{\bar{\cal I}}}\left[\cos\left(\Delta_{i}\right)-\frac{P^{(N)}-a_i^2}{2a_i\sqrt{NP^{(N)}}}\right],$
where $\Delta_{i}$ is the relative phase between $h_i$ and $z^{(N)}$;
\IF{$\cos\left(\Delta_{i^*}\right)>\frac{P^{(N)}-a_{i^*}^2}{2a_{i^*}\sqrt{NP^{(N)}}}$}
\STATE 1. Update
$z^{(N+1)}=\frac{1}{\sqrt{N+1}}\left(\sqrt{N}z^{(N)}+a_{i^*}e^{j\phi_{i^*}}\right)$ and $P^{(N+1)}=\left|z^{(N+1)}\right|^2;$
\STATE 2. Set ${\cal I}={\cal I}\cup i^*$ while excluding $i^*$ from $\bar{\cal I}$;
\ELSE \STATE Terminate the algorithm;
\ENDIF
\ENDFOR
\end{algorithmic}
\end{algorithm}

\section{Numerical Results}
This section presents numerical examples of the achievable performance of the proposed opportunistic collaborative beamforming with respect to the bounds developed in Section~\ref{sec:threshold} and the single-best-relay selection scheme proposed in \cite{Bletsas06}. All of the results in this section assume i.i.d.~channel coefficients $h_k = a_k e^{j \phi_k}$, $k\in\{1,\dots,K\}$, with amplitudes $a_k$ Rayleigh distributed with mean $\text{E}[a_k] = \frac{\sqrt{\pi}}{2}$ and phases $\phi_k$ uniformly distributed on $\left(-\pi,\pi\right]$.

To obtain numerical results for finite values of $K$, minor modifications were made to the ideal coherent upper bound and sector-based lower bound selection rules. These selection rules were developed for the case when $K\rightarrow\infty$ and are based on the statistics of the channel coefficients, not the current channel realization. Hence, when $K$ is finite, it is possible that no nodes meet the selection criteria. It is also possible that one or more nodes meet the selection criteria but the resulting power is less than that of the single best node. The modified ideal coherent upper bound and sector-based lower bound selection rules check for these cases and select the single best node if either case occurs.

\begin{figure}[htp]
\begin{center}
\includegraphics[scale=0.43]{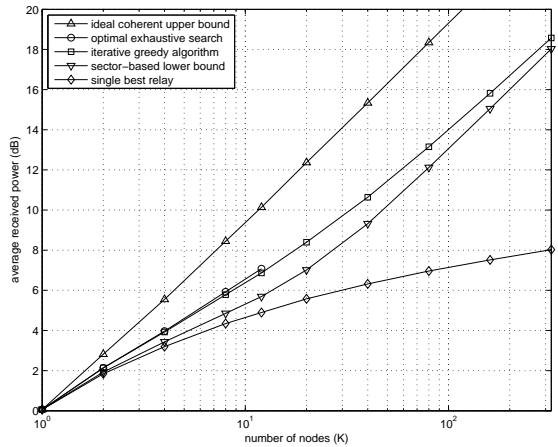}
\caption{Average received power versus the total number of nodes $K$.}\label{fig:receivedpower}
\end{center}
\end{figure}

Figure~\ref{fig:receivedpower} shows the average received power as a function of the total number of nodes $K$. The optimum opportunistic collaborative beamformer performance is plotted only for $K \le 12$ due to the computational complexity of the exhaustive search over $2^K-1$ possible selection vectors. The upper and lower bounds confirm that the received power scaling of opportunistic collaborative beamforming is linear in $K$ and, as predicted in (\ref{eq:bothbounds}), their performance gap is approximately 6.37dB for large~$K$. These results also demonstrate that the iterative greedy algorithm outperforms the sector-based selection algorithm and exhibits an average received power performance very close to the optimum exhaustive search, at least for $K \le 12$, with much lower computational complexity.

%%%%%%%%%%%%%%%%%%%%%%%%%%%%%%%%%%%%%%%%%%%%%%%%%%%%%%%%%%%%%%%%
%%%%%%%%%%%%%%%%%%%%%%%%%%%%%%%%%%%%%%%%%%%%%%%%%%%%%%%%%%%%%%%%

\begin{figure}[htp]
\begin{center}
\includegraphics[scale=0.43]{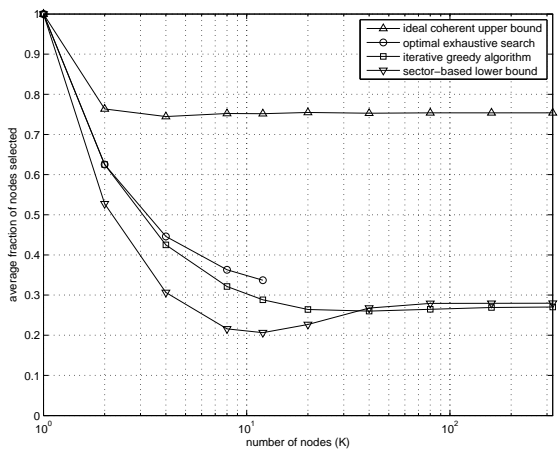}
\caption{Average fraction of nodes selected for participation in the collaborative beamformer versus the total number of nodes $K$.}\label{fig:selectionfraction}
\end{center}
\end{figure}

Figure~\ref{fig:selectionfraction} shows the average fraction of nodes selected for participation in the opportunistic collaborative beamformer versus the total number of nodes $K$. In the case of the ideal coherent upper bound, the fraction of nodes selected converges to about $75\%$, which agrees well with our analytical result $\Pr\left(a_k \ge r^*\right) = e^{-0.5316^2}\approx 0.7538$. This can be further explained by the fact that the nodes all have identical phase and only nodes with insufficient amplitude are rejected. For $K \le 12$, the optimum exhaustive search selection rule tends to be more inclusive than either the iterative greedy algorithm or the sector-based selection algorithm. For large $K$, the iterative greedy algorithm and the sector-based selection rule tend to select similar fractions of nodes for beamforming, with the sector-based selection being slightly more inclusive in this scenario.

%%%%%%%%%%%%%%%%%%%%%%%%%%%%%%%%%%%%%%%%%%%%%%%%%%%%%%%%%%%%%%%%
%%%%%%%%%%%%%%%%%%%%%%%%%%%%%%%%%%%%%%%%%%%%%%%%%%%%%%%%%%%%%%%%

\section{Discussion and Conclusions}

One of the appeals of opportunistic collaborative beamforming is that each node in the system requires only one bit of feedback in order to commence or halt transmission. This is in contrast to fully-coherent collaborative beamforming schemes that typically require several bits of feedback per node in order to perform local phase pre-compensation (and perhaps additional bits to exclude nodes with weak channels from transmitting). The rate at which the source selection vectors must be sent depends on the channel coherence time as well as the relative frequencies of the nodes' local oscillators. In systems with channels that exhibit long coherence times, feedback will be required at a rate inversely proportional to the maximum carrier frequency difference among the nodes. Outlier nodes with large carrier offsets could be permanently excluded from the pool of available nodes to reduce the feedback rate requirement. More detailed studies on the feedback rate requirement for opportunistic collaborative beamforming under general channel conditions are of importance.

Throughout our previous discussions, we have concentrated on the centralized selection in which the BS feedbacks the selection decision to the source nodes. However, it is worth emphasizing that the threshold-based selection algorithm can be also easily implemented in a distributed manner. We assume that each  node only has perfect knowledge about its own channel by exploiting a pilot signal transmitted from the BS. Similar to \cite{Bletsas06}, we can consider a system where each node sets a timer inversely proportional to its channel gain. Upon its timeout, the node with the strongest channel gain first broadcasts its own channel information (amplitude and phase) to its peer nodes. This is in contrast to \cite{Bletsas06} in which the best node simply starts sending data to the BS. Exploiting the received information about the strongest channel gain, each node can compare its own channel amplitude and phase against some pre-designed thresholds. In the next time slot, the nodes with channel conditions exceeding the thresholds start transmitting data simultaneously with the best node.

The main contributions of this work are the development of an energy-efficient opportunistic collaborative beamformer with one-bit feedback and a unification of the ideas of collaborative beamforming and relay selection. Unlike conventional collaborative beamforming, opportunistic collaborative beamforming is applicable in networks with nodes that may not be able to control their carrier frequency or phase. While optimal node selection for opportunistic collaborative beamforming is exponentially complex in the number of available nodes, we showed that low-complexity selection rules can provide near-optimum beamforming gain with performance within 6.37dB of an ideal fully-coherent collaborative beamformer. We also showed, in contrast to single-best-relay selection, that the received power of opportunistic collaborative beamforming scales linearly with the number of available nodes.

\bibliographystyle{IEEEtran}
\bibliography{Bib}
\end{document}